%% file: main.tex
\RequirePackage{rotating}
\documentclass[sigconf,nonacm]{acmart}

\usepackage{soul}
\usepackage{tabularx}
\usepackage{colortbl}
\usepackage{multicol}
\usepackage{enumitem}
\usepackage{rotating}

\newcommand{\coo}{\ensuremath{\mathrm{CO_{2}e}}}

\AtBeginDocument{%
  }

\begin{document}

\title[The HCI GenAI $CO_2$ST Calculator]{The HCI GenAI $CO_2$ST Calculator: A Tool for Calculating the Carbon Footprint of Generative AI Use in Human-Computer Interaction Research}

\author{Nanna Inie}
\email{nans@itu.dk}
\orcid{0000-0002-5375-9542}
\affiliation{%
  \institution{IT University of Copenhagen, Center for Computing Education Research}
  % \streetaddress{Rued Langgaards Vej 7}
  \city{Copenhagen}
  \country{Denmark}
}

\author{Jeanette Falk}
\email{jfo@cs.aau.dk}
\orcid{0000-0001-7278-9344}
\affiliation{%
  \institution{Aalborg University, Copenhagen, Department of Computer Science}
  % \streetaddress{A. C. Meyers Vænge 15}
  \city{Copenhagen}
  \country{Denmark}}

\author{Raghavendra Selvan}
\email{raghav@di.ku.dk}
\orcid{0000-0003-4302-0207}
\affiliation{%
  \institution{University of Copenhagen, Department of Computer Science}
  % \streetaddress{Universitetsparken 1}
  \city{Copenhagen}
  \country{Denmark}
}

\renewcommand{\shortauthors}{Inie et al.}

\begin{abstract}
  Increased usage of generative AI (GenAI) in Human-Computer Interaction (HCI) research induces a climate impact from carbon emissions due to energy consumption of the hardware used to develop and run GenAI models and systems. The exact energy usage and and subsequent carbon emissions are difficult to estimate in HCI research because HCI researchers most often use cloud-based services where the hardware and its energy consumption are hidden from plain view. The HCI GenAI $CO_2$ST Calculator is a tool designed specifically for the HCI research pipeline, to help researchers estimate the energy consumption and carbon footprint of using generative AI in their research, either a priori (allowing for mitigation strategies or experimental redesign) or post hoc (allowing for transparent documentation of carbon footprint in written reports of the research). 
\end{abstract}




\begin{teaserfigure}
\centering
  {%
    \includegraphics[height=7cm]{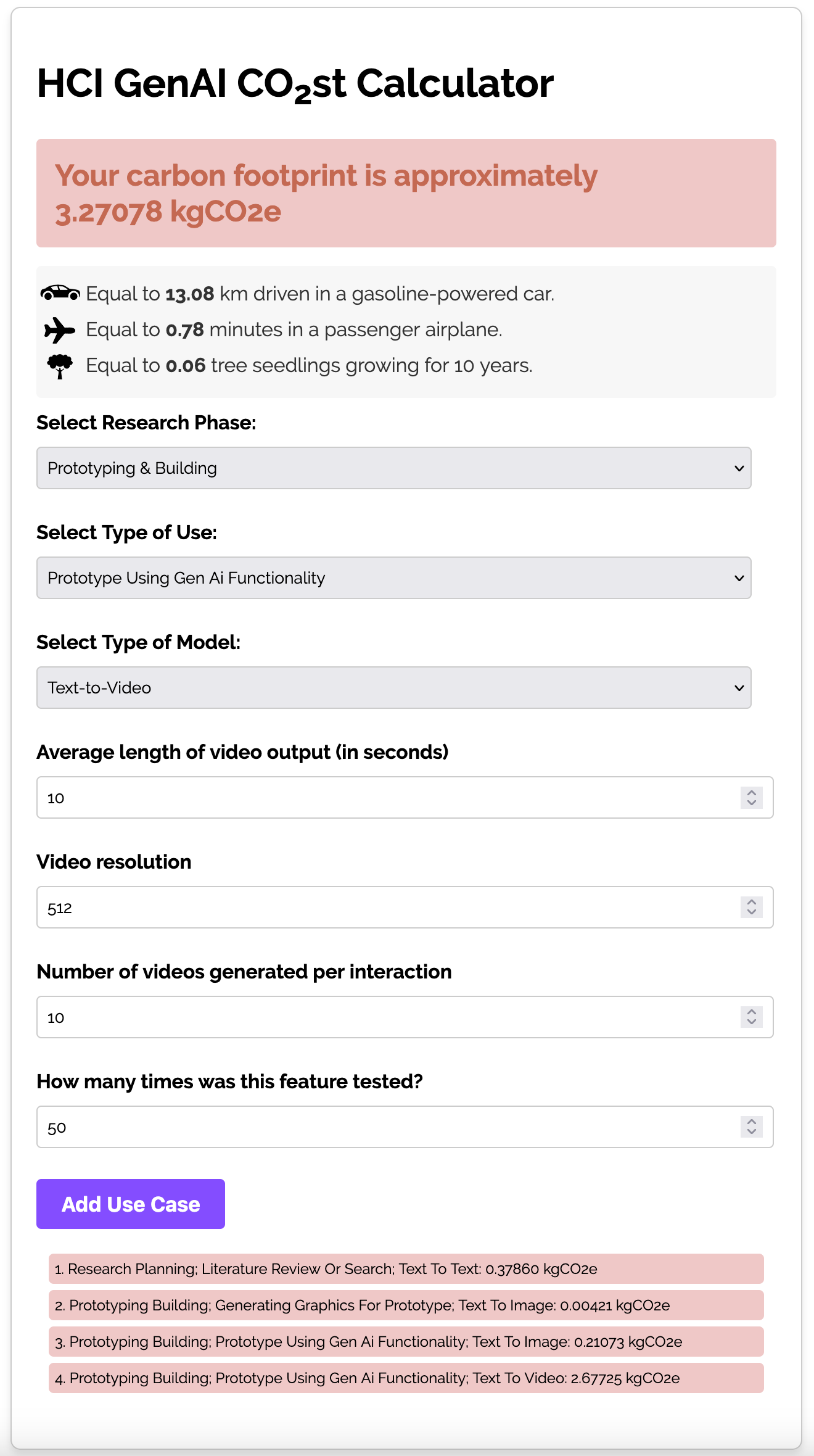}
    }
    $\qquad$
    {%           
    \includegraphics[height=7cm]{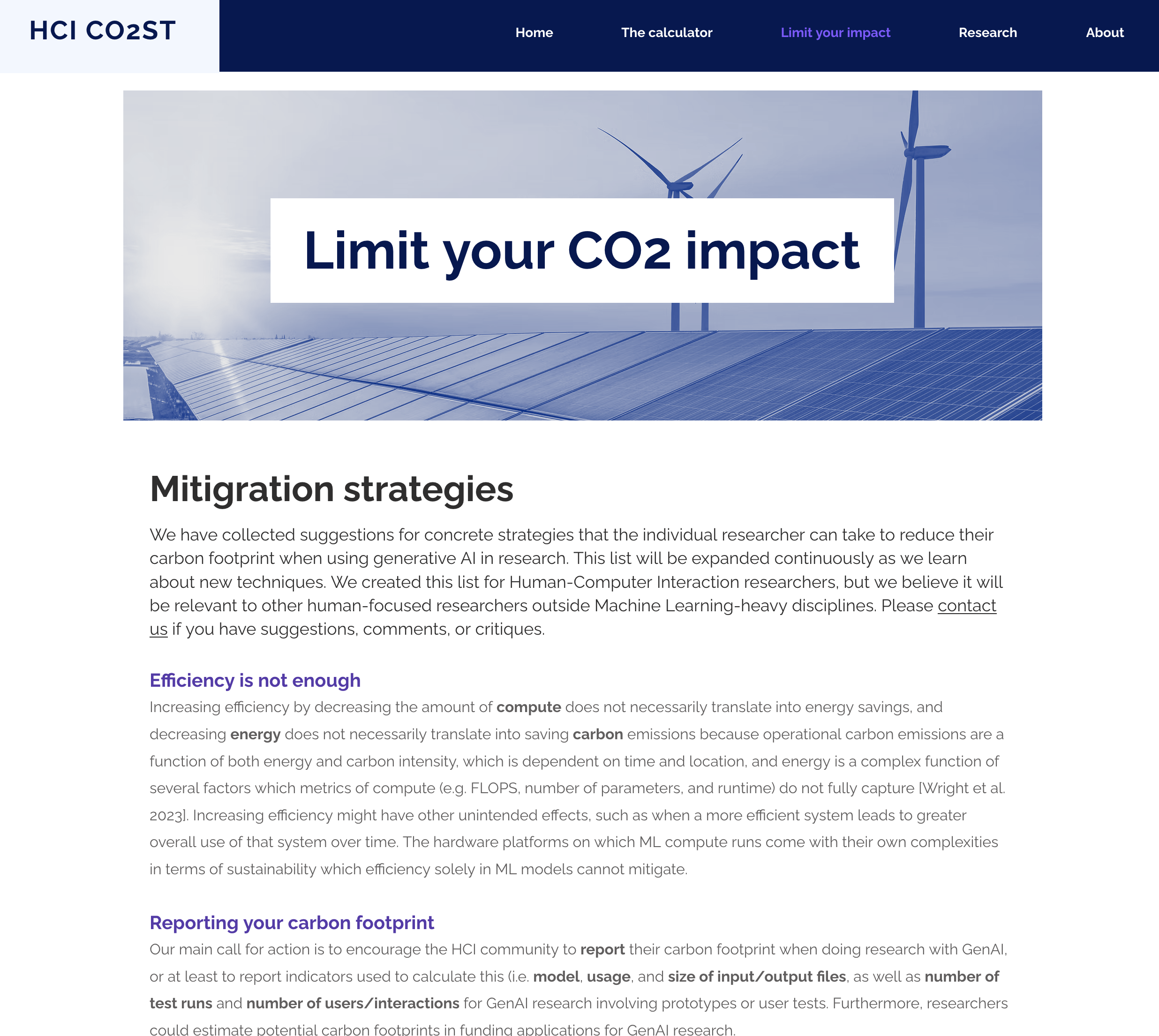}
    }
  \caption{Left: A screenshot of the HCI GenAI $CO_2$ST calculator. Right: A screenshot of the website offering mitigation strategies for HCI researchers who wish to limit their carbon footprint when doing research with generative AI.}
  %\Description{Left: A screenshot of the HCI GenAI $CO_2$ST calculator. Right: A sketch of the proposed interaction. }
  \label{fig:teaser}
\end{teaserfigure}

\received{20 February 2007}
\received[revised]{12 March 2009}
\received[accepted]{5 June 2009}

\maketitle

\vspace{-1em}
\section{Introduction and motivation}
%Should contain: "A description of the relevance of the work to the immediate CHI conference community, as well as to the broader CHI community, emphasizing its novelty, uniqueness, and rationale."}

The extensive use of generative AI (GenAI) models in research worldwide has a significant carbon footprint. In recent years, the electricity use by Meta, Amazon, Microsoft, and Google --- main providers of cloud compute services --- has more than doubled, and the electricity consumption by global data centers has increased by 20-40\% \cite{luccioni2024power, hintemann2022cloud}. Irish data centers are on the path to derail the entire country's climate targets \cite{theguardianIrelandsDatacentres}. 

In this paper we present the \textit{HCI GenAI $CO_2$ST Calculator}: a calculator designed specifically for Human-Computer Interaction (HCI) researchers to estimate the carbon footprint due to electricity consumption of GenAI use in their research. When researchers explore, test, and prototype with GenAI, their use causes additional $\coo$ consumption. Additionally the downstream effects of GenAI being integrated into more systems, processes, and user interactions because of our research is something that the HCI community have a share in responsibility for. 

With this calculator, we hope to make two contributions to the HCI community: First, we wish to enable HCI researchers to be fully transparent of their own research and acknowledge their own climate impact. Second, we hope to evoke critical thinking about the importance and necessity of HCI research conducted with the use of GenAI.

\section{Background --- sustainable HCI and carbon tracking}

In 2007, Blevis coined the term Sustainable Interaction Design (SID) and argued that ``sustainability can and should be a central focus of interaction design'' \cite{blevis2007sustainableIXD}. This perspective includes the responsible audit of tools we use to conduct our research. The large-scale adoption of GenAI tools is more than likely to contribute to the replication of ``our modern society’s overconsumption habits of natural resources within the digital space'' \cite{utz2023climate}.

Research on sustainable AI and ML generally fall into two camps: AI and ML \textit{for} sustainability and sustainability \textit{of} AI and ML, see e.g. \cite{sustainableaiAboutBonn, wright2023efficiencyenoughcriticalperspective}. While a growing number of publications are directed towards AI \textit{for} the UN Sustainable Development Goals, there is little research addressing the, often hidden, environmental costs of AI \cite{sustainableaiAboutBonn}. These efforts are significantly higher in ML and AI communities than in HCI. For example, because a model's architecture can affect how much power it consumes \cite{derczynski2020power}, different more energy-efficient approaches in the IT-infrastructure, data, modeling, training, deployment, and evaluation of ML models have been suggested --- see \citet{mehlin2023energyefficientdeeplearningoverview,bartoldson2023compute}.

Assessing the full climate impact of computing is complex: Environmental impact due to mining for rare earth minerals, green house gas (GHG) emissions due to hardware manufacturing, pollution due to e-waste disposal, and water consumption for cooling are all contributing factors~\cite{strubell2020energy}. In this work, we focus on \textit{energy consumption} during development and deployment of GenAI models, which constitutes a substantial share of the overall climate impact of GenAI~\cite{henderson2020towards,anthony2020carbontracker}, and which is a factor that can be mitigated by the individual researcher.

There are several carbon and energy tracking tools available --- Wright and colleagues discuss pros and cons of seven of these \cite{wright2023efficiencyenoughcriticalperspective}. None of them, however, have the focus of HCI research, and many of their metrics do not make sense in an HCI context (such as choosing the type of hardware used for the computation and the ML tasks performed). Similarly, mitigation strategies directed at the architecture and training of models are rarely relevant to researchers outside ML, who rely on off-the-shelf, \textit{multi-purpose} models. This exacerbates the sustainability issue for HCI researchers, since multi-purpose, generative architectures (such as the GPT models) are \textit{orders of magnitude} more expensive than task-specific systems \cite{luccioni2024power}. The lack of transparency from large multi-purpose model providers (such as OpenAI, Microsoft, and Google) about critical data, such as model training and hosting, complicates the issue.

\section{The HCI $CO_2$ST Calculator}

The HCI $CO_2$ST Calculator is, at its core, a calculator through which the researcher can input how they used GenAI in their research, which model they used, how much they used it, and the calculator will give an estimate of the $\coo$ in kilos, this GenAI research use has cost. The estimate is based on a combination of energy measurements from experiments conducted with publicly available models run on our own hardware (see section \ref{sec:backend}), and a thorough review of how GenAI was used in submissions from the CHI 2024 corpus which used GenAI in some form \cite{iniefalk2025co2st}. In contrast to existing carbon footprint calculators\footnote{E.g., \url{https://mlco2.github.io/impact/} \newline \url{https://www.deloitte.com/uk/en/services/consulting/content/ai-carbon-footprint-calculator.html} \newline \url{https://www.carbonfootprint.com/calculator.aspx \newline https://genai-impact.org/blog/post-2/}}, which tend to focus on technical features such as hardware, ML tasks, and world location where the training was performed, the HCI calculator is designed to match the level of detail likely to be relevant to HCI researchers. Since we can not know the exact training and running cost of many of the models, we emphasize that all results are \textit{estimates} and that calculations err on the conservative side.

\subsection{Which information is necessary?}

The factors we use to calculate a credible estimate of $\coo$ consumption of model use are: \textbf{\textit{model type}}, \textbf{\textit{usage numbers}}, and \textbf{\textit{input/output resolution}} (depending on the model type). If GenAI is integrated into a prototype or used as part of a user study, we need to know the \textbf{\textit{number of test runs}} and \textbf{\textit{number of interactions}} with the system) --- an overview is shown in Table \ref{tab:transparencyfactors}. However, not all of these factors are relevant to all HCI research pipelines. 

\renewcommand{\arraystretch}{1.4}
\begin{table*}
        \centering
        \caption{Factors necessary for calculating the energy usage and carbon footprint for each model use case. The HCI Transparency Factors represent factors which we, as HCI researchers, have the possibility to know and record for full GenAI use transparency.} 
        \scalebox{0.9}{
    \begin{tabularx}{0.8\textwidth}{lXr}
        \hline \hline
        \rowcolor[HTML]{EFEFEF} 
        \textbf{HCI Transparency Factor}                 & \textbf{Explanation} \\ \hline 
        \texttt{model}                  & Name or type of model (text-to-text, text-to-image, audio-to-text etc.) \\ 
        \texttt{usage \#}               & Number of prompts, images generated, minutes transcribed etc. \\ 
        \texttt{resolution}             & Length of prompts/size of dataset, resolution of images generated  \\ \hline
        
        \multicolumn{2}{l}{If the research involves the integration of GenAI into a prototype or system and/or user tests, further:} 
         \\
        \texttt{\# test runs}           & Number of tests of the system during prototyping or other factors which makes it possible to estimate how many times, the API has been called. \\ 
        \texttt{\# interactions}        & Number of interactions with the system during the evaluation. \\ \hline
    \end{tabularx}}
    \vspace{2mm}
    \Description{The table shows two columns entitled Factor and Explanation. The first three rows shows factors which should be considered if the research does not use GenAI as part of a system or prototype. These factors are: Model (this is the name or type of model, such as text-to-text, text-to-image, audio-to-text, etc), usage (this is the number of prompts, images generated, minutes transcribed etc), and resolution (this is the length of prompts or size of the dataset, and resolution of images generated). The last two rows show factors which should be considered if GenAI is integrated into a system. These are number of test runs during prototyping and number of interactions with the system during the evaluation.}
    \label{tab:transparencyfactors}
\end{table*}

The goal is to translate the technical factors affecting the carbon footprint of GenAI use into an interface that makes it easy for an HCI researcher to audit their empirical research. We hypothesize that the categorization imposed by the calculator will prompt reflection about different types uses that might incur $CO_2$st which the researcher had not thought of, such as automatic transcription, automatic proofreading, or the generation of images for slides for a conference presentation.

\subsection{Reflecting typical HCI research pipelines}

The input fields reflect typical pipelines of HCI research and, for simplicity, some of the inputs are based on averages. For example, when the user records a \textit{Literature review and search} under the Research planning phase, we do not expect them to count the exact about of characters or words of each of the articles that they have input into a given system. Instead, we assume an average of 6000 words per article, and perform the calculation based on that average. All of these estimates and averages are clarified on the website under \textit{Research}, and our categorization of GenAI use in different research phases is described in Table \ref{tab:hciphases}.

\begin{table*}
\renewcommand{\arraystretch}{1.2}
\caption{A seven-stage model of HCI research phases, adopted from Elagroudy and colleagues \cite{elagroudy2024transforming} and appropriated to reflect the observed use of GenAI in HCI research.} 
\scalebox{0.8}{
\begin{tabular}{|p{2.7cm}|p{2.7cm}|p{2.7cm}|p{2.7cm}|p{2.7cm}|p{2.7cm}|p{2.7cm}|}
\hline
\rowcolor[HTML]{EFEFEF} 
\textbf{Research \newline Planning} & \textbf{Prototyping \newline \& Building} & \textbf{Evaluation \newline \& User studies} & \textbf{Data Collection} & \textbf{Analysis \newline \& Synthesis} & \textbf{Dissemination \& Communication} & \textbf{Training \newline \& Fine-tuning} \\ \hline

\textbf{Description}: Where a research subject is selected and a research method is chosen. & \textbf{Description}: Where a product or system is built. & \textbf{Description}: Where a product is evaluated, used, or explored by users. & \textbf{Description}: Where the researcher gathers data for analysis. & \textbf{Description}: Where the researcher makes sense of the data. & \textbf{Description}: Where the results are communicated to the scientific community. & Where GenAI models are made or redesigned specifically for HCI research purposes. \\ \hline

\begin{flushleft}
\textbf{Example use:}
 \begin{itemize} [leftmargin=*]
    \item Identifying research gaps
    \item Generating study materials
    \item Literature search
    \item Study design
    \item CHI workshops and courses
 \end{itemize}
\end{flushleft}

& 
\begin{flushleft}
\textbf{Example use:}
 \begin{itemize}[leftmargin=*]
     \item Integrating GenAI functionality into prototypes.
     \item Using GenAI to generate code for systems.
     \item Generating content or visuals for a prototype, system, or probe.
 \end{itemize}  
\end{flushleft}
&
\begin{flushleft}
\textbf{Example use:}
\begin{itemize}[leftmargin=*]
    \item User evaluation.
    \item User studies.
\end{itemize}
\end{flushleft}
& 
\begin{flushleft}
\textbf{Example use:}
\begin{itemize}[leftmargin=*]
    \item Generating data for exploration of the output.
    \item Generating data for evaluation of the output.
    \item Transcription of audio data.
    \item Simulating human-generated data.
\end{itemize} 
\end{flushleft}
& 
\begin{flushleft}
\textbf{Example use:}
\begin{itemize}[leftmargin=*]
    \item Qualitative analysis.
    \item Quantitative analysis.
    \item Data trend identification.
\end{itemize} 
\end{flushleft}
&
\begin{flushleft}
\textbf{Example use:}
\begin{itemize}[leftmargin=*]
    \item Generation of manuscript text.
    \item Generating suggestions for text improvement.
    \item Generating graphics for articles and presentations.
\end{itemize}
\end{flushleft}
& 

\textbf{Example use:}
\begin{itemize}[leftmargin=*]
    \item Training novel GenAI models.
    \item Fine-tuning existing GenAI models.
\end{itemize} 

\\ \hline 

\end{tabular}}
\vspace{2mm}
\Description{The table shows a seven-stage model of HCI research with examples under each headlines. The columns are: 1. Research planning, which includes: identifying research gaps, generating study materials, literature search, and study design. 2. Prototyping and building, which includes: Integrating GenAI functionality into prototypes, using GenAI to generate code for systems, and generating content or visuals for a prototype, system, or probe. 3. Evaluation and user studies, which includes: Evaluation of prototypes with users, user studies
with off-the-shelf GenAI products (such as ChatGPT). 4. Data collection, which includes: Generating data for exploration of the output, generating data for evaluation of the output, transcription of audio data, simulating human-generated data. 5. Analysis and synthesis, which includes: Qualitative analysis, quantitative analysis, and data trend identification. 6. Dissemination and communication, which includes: Generating manuscript text, Generating suggestions for text improvement, and generating graphics for articles and presentations, and 7. Training and fine-tuning, which includes: Training novel GenAI models for HCI research, and fine-tuning existing GenAI models for HCI research.}
\label{tab:hciphases}
\end{table*}

\subsection{Front-end and design}

This section explains the front-end design of the calculator module only. For more about the online version, see Section \ref{sec:online}. Based on \citet{iniefalk2025co2st}, we create a flow that begins with choosing which Research phase the use was part of (Research planning, Prototyping \& building, Evaluation \& user studies, Data collection, Analysis \& synthesis, Dissemination \& communication, or AI model training or fine-tuning) (see Table \ref{tab:hciphases} in the appendix for more detail about the research phases). Based on a user's selection of research phase, the input fields will change to reflect which factors needs to be input to obtain an estimate.

\begin{figure*}
    \centering
     {%
    \includegraphics[height=8cm]{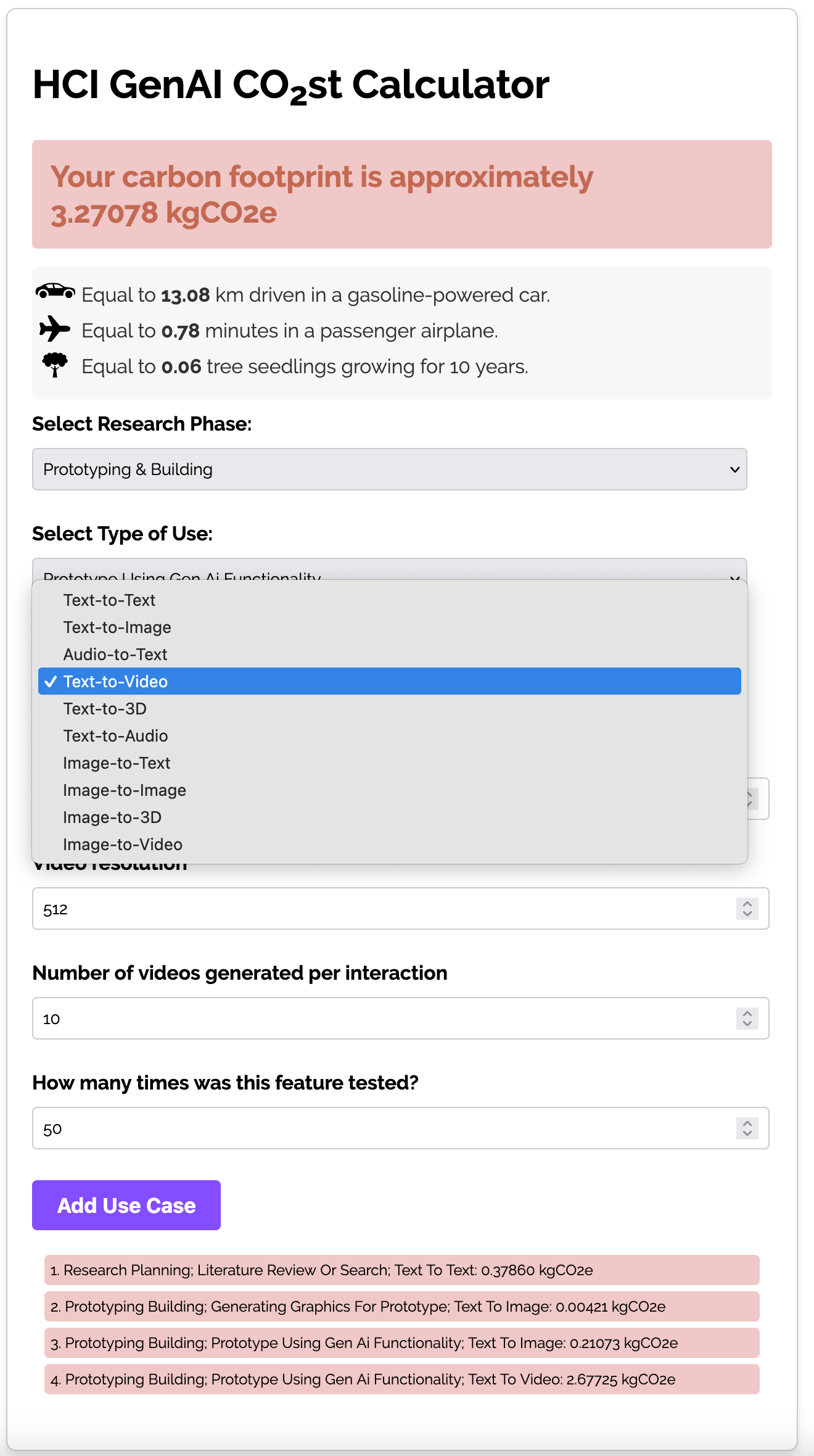}
    }
    $\qquad$
    {%           
    \includegraphics[height=8cm]{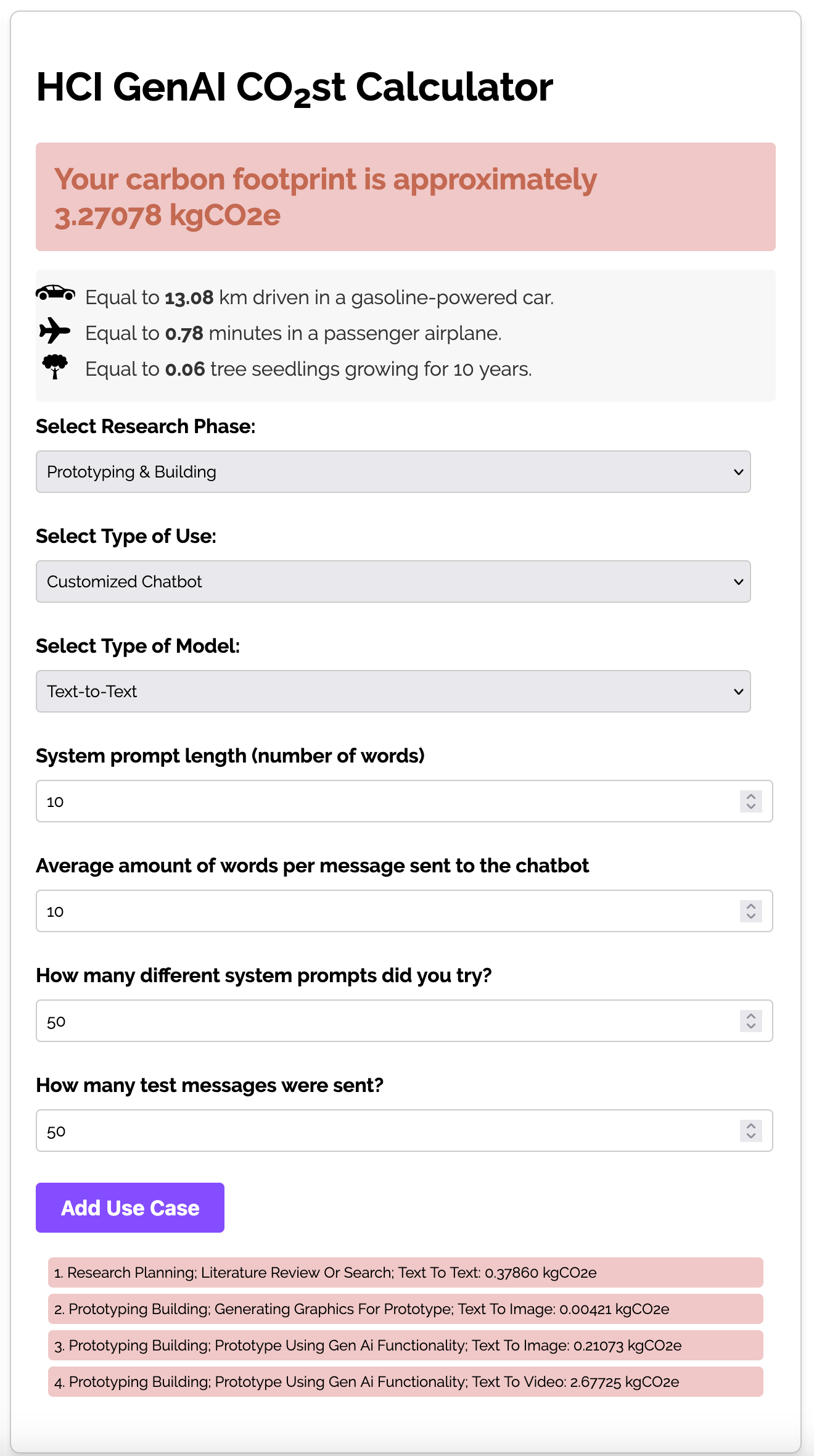}
    }
    \caption{Screenshots from the calculator, showing how the input fields change when the user chooses different research phases and ``stack'' their use cases to account for the entire research pipeline. For the Prototyping using GenAI functionality use type, the user can choose between all model types (left), while for the Customized chatbot type of use, the model types is ``locked'' to text-to-text (right).}
    \label{fig:interaction}
\end{figure*}

Figure \ref{fig:interaction} shows two examples from the calculator. We see that the input fields are different when the \textit{Type of use} is changed, mirroring the direct relevance to HCI research and simplifying the input. We have generally attempted to constrain the input fields to fewer, but more specific options in order to limit the amount of choices, the user has to make. We hope that, despite imposing constraints that are likely to miss some of the unique HCI research pipelines, the process of choosing between the available research processes and adding an individual use case per use encourages reflection on the extent of each research pipeline and the $CO_2$st it incurs. In the categorization we have maintained that GenAI use cases observed in the \textit{CHI 2024 corpus} are represented \cite{iniefalk2025co2st}. The list is obviously open to change and expansion. 

The result of the calculation is shown in a colored box on top of the page and updated when the user presses ``Add use case''. Use cases can be stacked because each research pipeline is likely to incur several GenAI uses, e.g., one for prototyping, and one for the subsequent user evaluation of a prototype, one for automatic transcription of audio data, and so forth. The result in kg$\coo$ is translated into equivalent numbers: km driven in a gasoline-powered car, number of minutes as a passenger on a commercial airplane, and number of tree seedlings grown for 10 years. These numbers are based on the EPA Greenhouse Gas Equivalencies Calculator. \footnote{\url{https://www.epa.gov/energy/greenhouse-gas-equivalencies-calculator}}

\subsection{Back-end and algorithms}
\label{sec:backend}

%The HCI GenAI $CO_2$ST calculator estimates the carbon footprint of different tasks based on the input provided by users. These numbers capture the usage; given this, the tool estimates the carbon footprint for different types of tasks based on measurements performed by the authors. 

At a high-level, for each task we have measured the energy consumption for a single use (or prompt) denoted $E_p$ watt-hour (kWh). Using an in-house set-up comprising an NVIDIA-RTX3090 GPU, Intel-i7 processor with 32GB memory, we measured the energy consumption for various models using Carbontracker~\cite{anthony2020carbontracker}, which are reported in Table~\ref{tab:energy}. The specific models shown in this table are used as proxies for the different model types (text-to-text, text-to-image, etc.) based on their popularity, ease-of-use, and availability (open-source). These choices provide useful approximations of the actual costs, which can vary between users due to differences in models and hardware used.
% \textcolor{red}{Raghav: One line about why these are reasonable proxies to use?}

We have not included multi-modal models in the calculator, but instead reduced them to the most computationally heavy parameters (which results in a conservative estimate). For example, text+image-to-image becomes image-to-image or image+video-to-image becomes video-to-image, and so forth. This is done partially for simplicity of the interface, and partially to reduce our own carbon footprint when reproducing experiments. 
% \textcolor{red}{Raghav: double-check this paragraph}.

We aggregate the usage information based on the user input into $N$, which is then used to estimate the overall energy consumption per use-case: $E=N\cdot E_p$ (kWh). This energy consumption is then converted to the carbon footprint using the global average carbon intensity of $CI = 0.481$ (kgCO2e/kWh)~\cite{ourworldindata}. The final carbon footprint $C$ is estimated as: $C = CI \cdot E$ (kgCO2e).

\input{models}

%\subsection{Conference version --- physical interaction}
%The HCI GenAI $CO_2$ST calculator is intended to be displayed in two versions: a physical exhibition and an online version. The physical exhibition of the calculator (see Figure \ref{fig:teaser}) is an interactive piece, integrating a touch display in the stem of a large cardboard tree which has balloons for leaves. The glowing ``leaves'' of the tree are white and soft pink to evoke associations with spring and the Sakura cherry trees of Japan. The tree is chosen as an exhibition piece to manifest the otherwise abstract relationship between computing use and its direct impact on climate and nature. Trees are one of our main sources for reducing $\coo$ in the atmosphere, and we hope the tree will inspire people to consider the value of natural resources consumed in the pipeline that feeds our technological ecosystem.

%A person can walk up to the tree and input how they have used GenAI in research --- we will suggest that they take departure in research conducted for CHI 2025 if they have submitted work to the conference --- and the screen will show an estimate of the $\coo$ that this research has cost. The user can then press a button to print their result, and a small thermal printer (also integrated into the stem of the tree) will print two receipts: one to hang on the tree, and one to bring home). The receipts hanging on the tree will constitute a growing body of evidence for the $CO_2$ST of GenAI use for HCI research. 

\subsection{Online version}
\label{sec:online}
The calculator is published on \textbf{\url{www.hcico2st.com}}. This website will be updated with relevant research and updated estimates as our knowledge about large GenAI models and their use in HCI expands. As more research pipelines are input, we can improve the estimates of \textit{average} carbon footprint for different research phases, allowing researchers to obtain post hoc estimates of their research even if they have not logged their detailed use of the models. Hosting the calculator on a website allows us to clarify and explain all data on which we base the calculations, as well as to present concrete mitigation strategies (\url{https://www.hcico2st.com/limit-your-impact}).

\section{Impact: awareness, transparency, and mitigation}

When planning research with GenAI there is a range of trade-offs which the individual HCI researcher can make to reduce their carbon footprint. Many of these are opaque to a user of cloud-based models, as the factors which increase $CO_2$ST are not clear or open. We intend for this system to have two practical impacts: First, to raise \textit{awareness} of the carbon footprint caused by GenAI as it is typically used in HCI research, and second, enabling the HCI community to expect and increase \textit{transparency} in reporting of research carbon footprint. The calculator will enable HCI researchers to report the estimated carbon footprint of their research in a research paper's ethical statement. Hopefully, both awareness and transparency will lead to increased reflection upon researchers' own practices and potentially mitigation strategies for the planning of future experiments.

The calculator will show that the energy consumption grows almost linearly with the task load i.e., longer prompts or more images or images of higher resolution cost more in energy. It will show that the far most carbon intensive research uses are training and fine-tuning new models and large-scale open-ended generation of datasets. Experiment designers could consider if increasing the task load to high ranges is always necessary. Prompting techniques can also be refined, reducing the need for several attempts. For research where users have to interact with GenAI models, users could be taught strategies for prompt engineering tailored for the specific research goal to reduce the amount of useless output. %The description of concrete mitigation strategies are shared with the online version of the calculator, and we hope the physical exhibition will spark discussion and downstream use of this research tool. 

\section{Summary}
This paper presents the HCI GenAI $CO_2$ST Calculator, a system designed to help HCI researchers estimate the carbon footprint of using generative AI in their research. The interface is designed to represent typical HCI pipelines, and the calculations performed by the calculator are based on estimates derived from experiments run on our own hardware. The calculator is intended to support HCI researchers in daily research practices. With this system, we hope to promote increased awareness and transparency in the HCI community about the climate impact of using GenAI in research.

\begin{acks}
Funding acknowledgments: NI received funding from the VILLUM Foundation, grant 37176 (ATTiKA: Adaptive Tools for Technical Knowledge Acquisition).
RS acknowledges funding received under Independent Research Fund Denmark (DFF) under grant 4307-00143B, and European Union’s Horizon Europe Research and Innovation programme under grant agreements No. 101070284, No. 10107040 and No. 101189771.
\end{acks}

\bibliographystyle{ACM-Reference-Format}
\bibliography{references}

\end{document}

%% file: models.tex
\begin{table}[]
\small
\begin{tabular}{llr}
\toprule
            {\bf Task} &            {\bf Model} &   $E_p$ (Wh) \\
\midrule
    Text-to-text &            Llama-3.1-Instruct~\cite{dubey2024llama} &  0.004685 \\
   Text-to-image & Stable-diffusion-XL~\cite{podell2024sdxl} &  0.001301 \\
   Audio-to-text &          Whisper~\cite{radford2023robust} &  0.006335 \\
   Text-to-Video &      AnimateDiff~\cite{lin2024animatediff} &  0.021742 \\
Text-to-3D model &           Shap-E~\cite{jun2023shap} &  0.026320 \\
   Text-to-Audio &         MusicGen~\cite{copet2024simple} &  0.011418 \\
   Image-to-text &             BLIP~\cite{li2022blip} &  0.003423 \\
  Image-to-image & Instruct-Pix2Pix~\cite{brooks2023instructpix2pix} &  0.000885 \\
     Image-to-3D & One-2-3-45~\cite{liu2023one2345} & 0.013010 \\
   Video-to-text &            XCLIP~\cite{ni2022expanding} &   0.001040 \\
  Video-to-video & RIFE~\cite{huang2022real}  &   0.026020 \\
  Audio-to-audio &          FreeVC~\cite{li2023freevc} &   0.006335 \\
  Image-to-video & SadTalker~\cite{zhang2023sadtalker} &   0.026020 \\
\bottomrule
\end{tabular}
\caption{Energy consumption per interaction for different model types.}
\label{tab:energy}
\end{table}